# Methodology for In-flight Flat-field Calibration of the Lyman-alpha Solar Telescope (LST)

Jing-Wei Li[1,2], Hui Li[1,2], Ying Li[1,2], Li Feng[1,2], Yu Huang[1,2], Jie Zhao[1,2], Lei Lu[1,2], Bei-Li Ying[1,2], Jian-Chao Xue[1,2]

[1] Key Laboratory of Dark Matter and Space Astronomy, Purple Mountain Observatory, Chinese Academy of Sciences, 10 yuanhua Road, Nanjing 210023, China; nj.lijw@pmo.ac.cn; nj.lihui@pmo.ac.cn;

[2] School of Astronomy and Space Science, University of Science and Technology of China, Hefei 230035, China

**ABSTRACT**

Flat-field reflects the non-uniformity of the photometric response at the focal plane of an instrument, which uses digital image sensors, such as Charge Coupled Device (CCD) and Complementary Metal-Oxide-Semiconductor (CMOS). This non-uniformity must corrected before being used for scientific research. In this paper, we assess various candidate methods via simulation using available data so as to figure the in-flight flat-field calibration methodology for the Lyman-alpha Solar Telescope (LST). LST is one of the payloads for the Advanced Space-based Solar Observatory (ASO-S) mission and consists of three instruments: a White-light Solar Telescope (WST), a Solar Disk Imager (SDI) and a dual-waveband Solar Corona Imager (SCI). In our simulations, data from the Helioseismic and Magnetic Imager (HMI) and Atmospheric Imaging Assembly (AIA) onboard the Solar Dynamics Observatory (SDO) mission are used. Our results show that the normal KLL method is appropriate for in-flight flat-field calibration of WST and implementing a transmissive diffuser is applicable for SCI. For the in-flight flat-field calibration of SDI, we recommend the KLL method with off-pointing images with defocused resolution of around 18", and use the local correlation tracking (LCT) algorithm instead of limb-fitting to determine the relative dis-placements between different images.

## 1. INTRODUCTION

Flat-field represents the inhomogeneous response to incident light of digitized detectors, such as Charge Coupled Device (CCD) and Complementary Metal-Oxide-Semiconductor (CMOS) detectors. For a telescope, flat-field may also include the non-uniformity of its optical components and it is called system-wide flat-field. The recorded images of a telescope must be corrected for the flat-field to obtain images more close to the target. Flat-field calibration is crucial to both ground-based and space-borne solar telescopes/instruments and sometimes even determines the usability of observational data.

The common flat-field calibrations for telescope using CCD and CMOS as detectors can be done via two approaches: (1) A uniform light source is used to illuminate the telescope or detecting system, and one or more images are collected to obtain a flat-field; (2) A stable light source is used to illuminate the telescope or detecting system to collect multiple images within the linear range of the detector's response with different exposure times. Linear fitting is applied to the rec-orded images with different exposure times and subsequently the obtained coefficients of all pixels form the demanded flat-field image. It should be noted that the flat-field obtained by illuminated telescope at or in front of its entrance is the system-wide flat-field, and that by illuminating detecting system is the flat-field just for detecting system (detector flat-field). And the former can reflect the imaging uniformity of the whole optical system on the detector.

For solar telescopes, it is often difficult to obtain a uniform or stable light source. Kuhn, Lin, Loranz (1991) proposed the so-called KLL method to obtain the flat-field of a solar telescope, which uses relatively stable non-uniform images as input, and was improved by Chae (2004a). The KLL method takes the Sun itself as the light source, and records multiple images of solar disk with different parts of the detector by changing the telescope's pointing. The obtained images are used to calculate the flat-field by iteration with some specific algorithm. In this method, the center and radius of the Sun in the recorded images are usually determined by fitting the solar limb, so as to derive the relative displacement between different images. At present, the KLL method is widely used by both ground-based and space-borne solar telescopes for full disk observation, such as the Optical and Near-infrared Solar Eruption Tracer (ONSET; Fang et al. 2013), the telescope for H-alpha full-disk observations of the Sun from Big Bear Solar Observatory (BBSO; Denker et al. 1999), the Helioseismic and Magnetic Imager (HMI; Schou et al. 2012) and the Atmospheric Imaging Assembly (AIA; Boerner et al. 2012; Lemen et al. 2012) onboard the Solar Dynamics Observatory (SDO) mission. Actually, the KLL method is also used for telescopes that take partial solar disk as their field-of-view (FOV), such as the Digital Vector Magnetograph attached to the 25-cm vacuum refractor at BBSO (Chae 2004b), and the Interface Region Imaging Spectrograph (IRIS; De Pontieu et al. 2014).

Other methods are proposed to obtain the flat-field for solar disk observation. Wang et al. (2017) pro-posed a flat-field calibration method based on ground glass. Wachter & Schou (2009) proposed a method for inferring small scale flat-field from the rotation of the Sun, while Bai et al. (2018) extracted the flat-field of a full-disk solar telescope from the routine observation data of one solar rotation. Furthermore, based on the Maximum Correntropy Criterion (MCC) method, Xu et al. (2016) got a flat-field for solar H-alpha observation with higher accuracy and faster convergence, and Gao et al. (2020) introduced image stitching technique to obtain a high-precision flat-field for instruments working at shorter wavelength.

For coronagraphs, the above flat-field calibration methods may not be applicable, because the key is-sue for observations to performable flat-field calibration is that the whole FOV of the detector should be illuminated with uniform or stable light source. Therefore, for most coronagraphs, a device is added to generate diffuse light so as to illuminate the corona-graph uniformly or steadily. For example, the Large Angle Spectroscopic Coronagraph (LASCO) onboard the Solar and Heliospheric Observatory (SOHO) has a diffuser window consisting of an opal and a neutral density filter, which is inserted into the aperture of the instrument to provide diffuse light when ready for flat-field observations (Brueckner et al. 1995). And the door of the Sun Earth Connection Coronal and Heliospheric Investigation (SECCHI) on the Solar Terrestrial Relation Observatory (STEREO) mission, which includes two traditional Lyot coronagraphs (COR1 and COR2), includes a diffuser that provides diffuse light for the detector when the door is closed (Howard et al. 2008). Bai et al. (2017) also proposed a flat-field measurement device and method for coronagraph based on opal glass, and demonstrated the feasibility of the method through simulation and observations.

In this paper, we present and assess various possible methods for the in-flight flat-field calibration of the Lyman-alpha Solar Telescope (LST; Li 2016; Li et al. 2019; Chen et al. 2019; Feng et al. 2019) for the Advanced Space-based Solar Observatory (ASO-S; Gan et al. 2015, 2019) so as to figure out the most plausible ones. We refer readers to next section for more information about the ASO-S mission and the LST payload. The paper is organized as follows: Section 2 introduces the ASO-S mission and the LST payload, and Section 3 presents the flat-field calibration methods for the White-light Solar Telescope (WST) of LST. In Section 4, we provide with emphasis and in much more details the methods for in-flight flat-field calibration of the Solar Disk Imager (SDI) of LST, because it is the most difficult one. We briefly introduce the method for the in-flight flat-field calibration of the Solar Corona Imager (SCI) of LST in Section 5, and finally present our discussions and conclusions in the last section (Section 6).

## 2. ASO-S MISSION AND LST PAYLOAD

The ASO-S mission is the first space-borne observatory approved by Chinese dedicated to solar observations (Gan et al. 2015, 2019) and scheduled to be launched in early 2022. The goal is to observe the Sun during the 25th solar maximum. The ASO-S mission has three payloads onboard: the Full-disk vector MagnetoGraph (FMG; Deng et al. 2019; Su et al.

2019), the LST, and the Hard X-ray Imager (HXI; Zhang et al. 2019). The scientific targets are the two large-scale eruptions on the Sun, i.e. solar flares and Coronal Mass Ejections (CMEs), and their associated solar magnetic fields.

LST includes a dual-waveband SCI, a WST, and a SDI. The FOV of both WST and SDI is 1.2 $R_\odot$ ($R_\odot$ stands for solar radius), and they operate at the 3600±20 Å and Lyα waveband (1216±75 Å), respectively. The FOV of SCI is 1.1-2.5 $R_\odot$, including polarized brightness measurement at 7000±400 Å (hereafter referred to as SCIWL) waveband and imaging at Lyα waveband (1216±100 Å) (hereafter referred to as SCIUV). For LST, we have four CMOS detectors, namely, WST detector, SDI detector, SCIWL detector and SCIUV detector. The intended CMOS detectors are manufactured by the GPixel Inc. in China. More details of the LST can be found in Li et al. (2019), Chen et al. (2019) and Feng et al. (2019).

In order to obtain science-grade data, in-flight calibration must be performed to ensure the data quality. Just like other space-borne solar imagers and coronagraphs, calibrations of the LST payload includes dark current, flat-field, radiometry, instrumental polarization, optical geometry, stray light, etc. The optical characteristics of the instrument and the response of detectors may change with time after launch. Therefore, in order to obtain more accurate observation data, it is necessary to carry out regular observations for the aforementioned calibrations. In this paper, we limited the scope to flat-field calibration. For different instruments or wavelengths, different flat-field calibration methods need to be adopted to meet the required accuracy of in-flight calibration.

According to the existing instruments scheme and scientific requirements, the demanded flat-field calibration accuracy of each LST instrument is 2%. LST flat-field calibration is scheduled weekly during satellite commissioning phase and every two to three months during normal satellite operation. It can also be adjusted on demand. In the following sections, we describe and assess the flat-field calibration methods for the three LST instruments, i.e., WST, SDI and SCI.

## 3.  ASO-S MISSION AND LST PAYLOAD

For the flat-field calibration of WST, we introduce the KLL method for system-wide flat-field and the light emitting diode (LED) method for detector flat-field.

### 3.1 KLL algorithm

Now that WST images the low solar atmosphere (photosphere) that in most cases changes slowly with time, the KLL algorithm could be a good candidate method for its in-flight flat-field calibration.

The KLL algorithm requires off-pointing images of the solar disk, i.e., imaging solar disk on different part of the detector with series of offset values. The selected offset values must meet two criteria: (1) the combined FOV of all the offset images must cover all pixels and be larger than the nominal FOV of WST, and (2) any pixels in the FOV of WST must be imaged at least twice with different offset values in order to improve the accuracy of the derived flat-field (Feng et al. 2019). These are also valid for the flat-field calibration by KLL method via satellite off-pointing (see Section 4). Li et al. (2020) has devised 13 and 21 offset locations, and here we adopt these locations and show them in Figure 1. The 13 offset locations are represented by the plus (+) sign and the 21 locations by plus (+) together with diamond (◊) signs. The values of their coordinates can also be found in Li et al. (2020).

The offset images of WST can only be obtained by pointing the satellite out of the Sun-spacecraft direction. It takes about 150 seconds for the satellite to change the pointing from one offset location to an adjacent one. WST can finish image acquisition for one location in 3 minutes. Li et al. (2020) demonstrated that the accuracy of derived flat-field of WST via KLL algorithm is not sensitive to time interval of image acquisition of adjacent offset locations (cf Figure 2). To be consistent with similar SDI image acquisition, we adopt 270 seconds as the time interval to estimate the accuracy of the derived flat-field with 21 offset values.

The Sun is an ideal source for the in-flight flat-field calibration with KLL algorithm for instruments that image the full disk photosphere, such as WST and HMI. Due to the lack of full disk images of the Sun in 3600 Å waveband from space, we use the intensity-grams of HMI at 6173 Å to estimate the possible accuracy of WST in-flight flat-field calibration. We take the released flat-field of HMI as a base flat-field and divide the derived flat-field by the base one to get a ratio image. The accuracy is defined as the ratio of the standard deviation to the mean value of the ratio image in the nominal

FOV of WST. From the flat-field data in Li et al. (2020) we infer that the estimated accuracy is approximately 0.5% – 0.6% (Figure 2).

## 3.2 LED Illumination

LED illumination is a backup method for WST in-flight flat-field calibration in case the above mentioned KLL algorithm is not applicable for any reasons. LED is generally a kind of stable light sources and for WST we need a stable UV LED, which has the required emission around 3600 Å waveband.

With the stable LED illumination, one takes images with different exposure times, which are chosen to guarantee that the detector has a linear response to the exposure times. The demanded high accuracy flat-field can be derived by linear fitting of the acquired images with different exposure times. The fitting gives slopes of all pixels and subsequently forms the slope image, which contains information of both the flat-field and the LED illumination (intensity), and the latter must be eliminated so as to get the flat-field. This can be done via dividing the slope image by the LED illumination/intensity image obtained during on-ground calibration or during the early commissioning phase of the satellite (in case there is some changes of the LED illumination during the launching process. We do not describe the method in more details because it is not the focus of this paper. It must be mentioned that if the LED does not illuminate all the optical components of WST, one cannot get the system-wide flat-field. For example, when the LED is just placed close to the detector, the inferred flat-field is just for the detector, which represents the non-uniform response of the detector pixels.

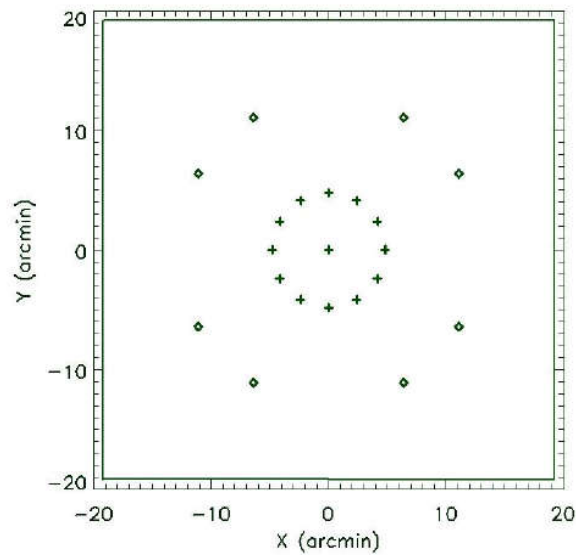

Fig. 1: The relative coordinates of the devised 13 and 21 offset locations with respect to the center of solar disk. The plus (+) sings represent the 13 offset locations, while the plus (+) together with diamond (◊) represent the 21 offset locations. The box stands for the full area of the detector (From Li et al. 2020).

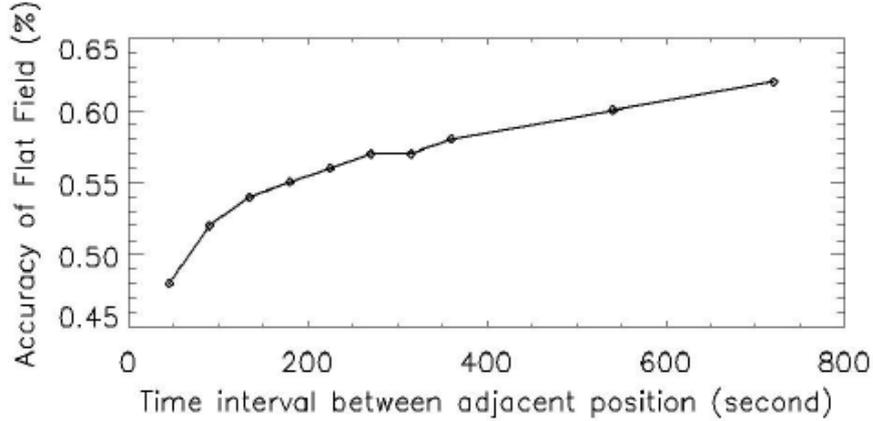

Fig. 2: Estimated accuracy of derived in-flight flat-field calibration of WST with time intervals between adjacent offset locations.

This method is also applicable for the in-flight flat-field calibration of SDI given the LED has the required emission in SDI's working waveband.

## 4. CALIBRATION METHODS FOR THE SDI INSTRUMENT

In this section, we describe various methods of the flat-field calibration based on the KLL algorithm and estimate the accuracy of the derived flat-fields in order to find the best one for the in-flight flat-field calibration of SDI.

### 4.1 Normal KLL methods

As mentioned above, the KLL algorithm requires images to be taken with a series of off-pointing values. As for the WST case, we adopt the proposed 13 and 21 offset values (see Figure 1) by Li et al. (2020) so that the flat-field calibrations of SDI and WST can be done simultaneously. With images at the 21 off-pointing locations one can get the flat-field of entire detector area including four corner regions while with that at the 13 off-pointing locations only the flat-field within the FOV of 1.2 $R_\odot$ can be derived.

After checking the images of the Sun in the Ly$\alpha$ waveband obtained by the Very high Angular resolution ULtraviolet Telescope (VAULT; Vourlidas et al. 2010) and images in the 304 Å and 1700 Å wavebands from AIA/SDO, we noticed that images in Ly$\alpha$ waveband are less structured than that in the 304 Å waveband and more structured than that in the 1700 Å waveband. Therefore, we use images in the 304 Å and 1700 Å wavebands from AIA to assess the feasibility of KLL algorithm for SDI flat-field calibration and the expected accuracy due to the lack of full disk images from space in the Ly$\alpha$ waveband.

Li et al. (2020) showed that shorter time interval between adjacent offset locations gives better accuracy of the derived flat-field and for SDI the time interval is better within 240 seconds, which is consistent with the ability of platform of ASO-S mission. Therefore, we fix the time interval to 240 seconds for the following simulations and estimations.

#### 4.1.1 SAT-KLL Method

We use SAT-KLL method to stand for the method based on the KLL algorithm and offset images are obtained via off-pointing the satellite. Here we introduce the data and procedure to mimic the satellite off-pointing. We downloaded 60 images in both the 304 Å and the 1700 Å wavebands obtained on 20 October 2010 from AIA data archive with reduced cadence of 240 seconds, and use them to simulate the required off-pointing images. We also downloaded the released flat-field of AIA in these two wavebands (Figure 3) and use them as the base (known) flat-field images for our analysis.

We choose 21 successively downloaded AIA images (we call them a set of images) to mimic the off-pointing images by shifting the downloaded images by the corresponding offset values. Then we multiply the shifted images by the down-loaded flat-field image, and use resultant images as the input images for the computation with the KLL algorithm. In this way we can generate 40 sets of images for input of KLL algorithm and subsequently get 40 derive flat-fields. More specifically, the first set consists of images 1 to 21, the second set consists of images 2 to 22, …, and the 40th set consists of images 40 to 60.

Similarly, when only considering the flat-field within the FOV of 1.2 $R_\odot$, we can choose 13 successive downloaded AIA images to form a set of images. Accordingly, the first set consists of images 1 to 13 while the 40th set consists of images 40 to 52.

With the above 40 sets of constructed off-pointing images and KLL algorithm, we derived 40 flat-field images for each of the 4 cases: 21 and 13 off-pointing images in both the 304 Å and the 1700 Å wavebands. One example of derived flat-field images in 304 Å and 1700 Å with 21 off-pointing images is shown in Figure 4.

**4.1.2 Estimation of Flat-field Accuracy**

We estimate the accuracy of the derived flat-field images by comparing them with the corresponding base flat-field image. Let $FC_i$ stands for the ith derived flat-field image, where i = 1, 2, 3, …, N and N is the total number of derived flat-field images, and $F_{base}$ for the corresponding base flat-field, we compute the ratio $R_i$ of the ith derived flat-field images to the base flat-field image via

$$R_i = \frac{FC_i}{F_{base}} \qquad (1)$$

We only concern the region within the SDI's FOV (i.e., 1.2 $R_\odot$), and hereafter we call it the region of interest (ROI). Figure 5 shows examples of computed ratios in the ROI. We calculate the standard deviation $\sigma_i$ and mean value $A_i$ of the $i$th ratio image in the ROI, and use the relative error $E_i$ as the accuracy of the $i$th derived flat-field image $FC_i$, that is

$$E_i = \frac{\sigma_i}{A_i} \times 100\% \qquad (2)$$

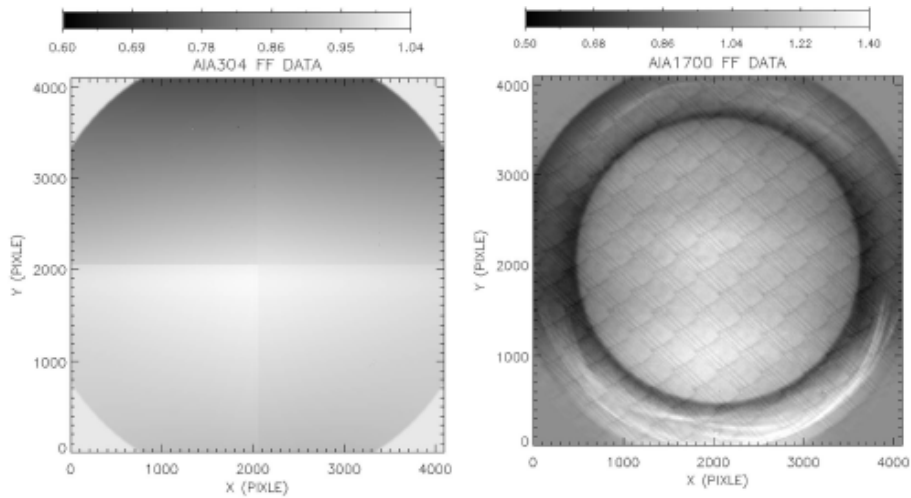

Fig. 3: The AIA flat-field in the 304 Å (left) and 1700 Å (right) wavebands obtained on 1 January 2012..

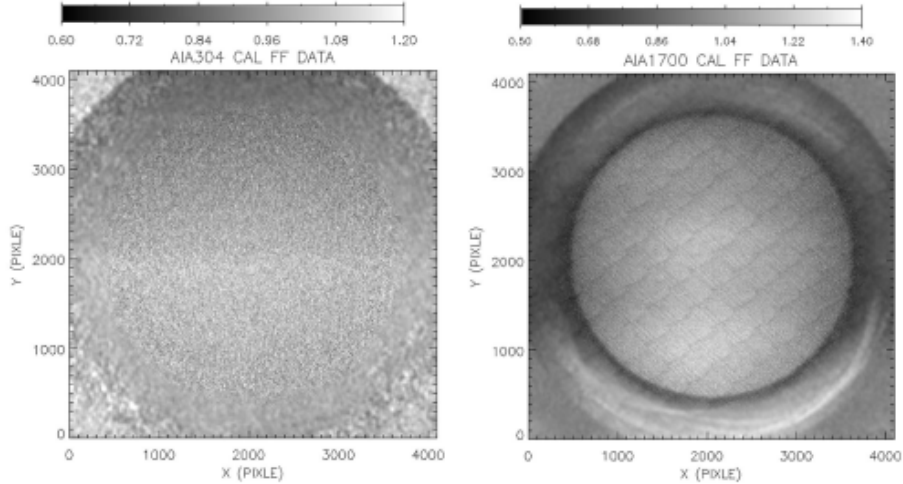

Fig. 4: One example of the 40 derived AIA flat-field images in 304 Å (left) and 1700 Å (right) wavebands with 21 off-pointing images and SAT-KLL method.

Then we calculate the average accuracy ($E_A$) of the derived flat-field images and take it as the possible flat-field calibration accuracy with the SAT-KLL method, namely

$$E_A = \sum_{i=1}^{N} E_i / N \qquad (3)$$

Now we use $E_{A304}$ and $E_{A1700}$ to represent the accuracy of the derived AIA flat-field images with above-mentioned SAT-KLL method in the 304 Å and 1700 Å wavebands, respectively. For the reason given in subsection 4.1, as an estimate, we take the average of $E_{A304}$ and $E_{A1700}$ as the possible accuracy ($E_{SDI}$) of SDI flat-field calibration with the SAT-KLL method, i.e.

$$E_{SDI} = (E_{A304} + E_{A1700})/2 \qquad (4)$$

Applying this process of accuracy estimation of SDI flat-field calibration with SAT-KLL method to the derived flat-field images in previous subsection (i.e., subsection 4.1.1), we get a possible accuracy of SDI flat-field calibration (with SAT-KLL method) of 5.91% and 8.13% for cases of 21 and 13 off-pointing images, respectively. The accuracy with 21 off-pointing images is better than that of 13 off-pointing images, but both are much worse than the required 2% accuracy. This may be because the KLL method itself requires relatively stable observation targets, and the time span of off-pointing images is large (about 80 minutes in the case of 21 images) and in this time span the Sun in the UV waveband may change significantly. Since using 21 off-pointing images gives better accuracy of flat-field, hereafter, we focus on the case with 21 off-pointing images.

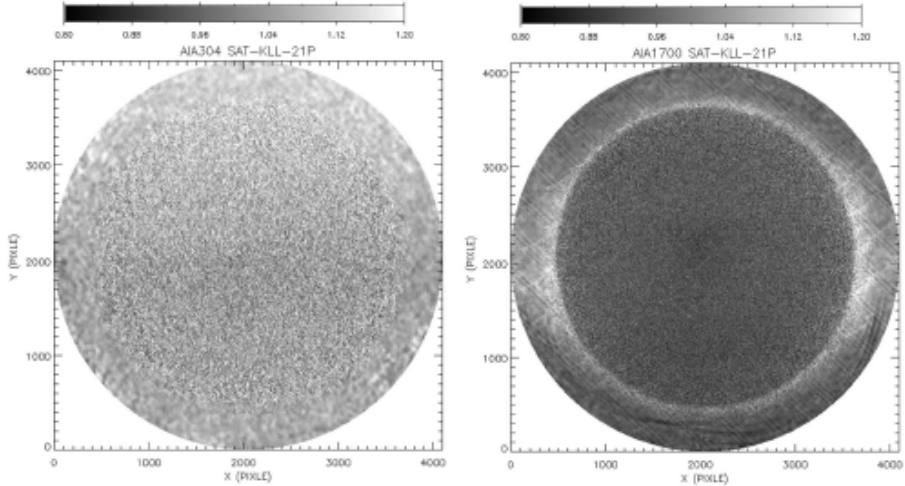

Fig. 5: Example of the ratio of the derived flat-field image to the base flat-field image of AIA within 1.2 $R_\odot$ FOV in the 304 Å (left) and 1700 Å (right) wavebands.

## 4.2 KLL Method with Image Defocusing

To reduce the influence of the changes of the Sun during the time span of image acquisition and the com-plicate structures of solar images in the UV waveband, we try to defocus the observed images used for the KLL algorithm. We call this method using defocused images DF-SAT-KLL method. The defocused images are obtained by using a two-dimensional (2D) Gaussian point spread function (PSF) to convolve the simulated off-pointing images and subsequently reduce the spatial resolution.

Gaussian PSFs with different full width at half maximum (FWHM) can be applied to the off-pointing images to get different reduced spatial resolutions (referred as defocused resolution for short). Bearing in mind that pixel resolution of AIA images is 0.6", by applying Gaussian PSFs with FWHM of 10, 20 and 30 pixels, we get defocused resolution of approximately 6", 12" and 18", respectively.

The KLL method is proposed basically for ground-based full-disk observation of the Sun in white-light, and the observation can be done in a relatively short time (a few minutes or less), for which the influence of solar rotation can be ignored. The common practice is to determine the coordinates of the center and radii of the Sun on the images by fitting the solar limb (referred as limb-fitting), so as to determine the relative displacement between different images, which are required for calculating the flat-field with KLL algorithm.

We defocus the off-pointing images obtained in subsection 4.1.1 to get images with defocused resolution of 6", 12" and 18" and use limb-fitting procedure to infer the displacement of different images. The defocused images for the reference position from AIA observations with 12" resolution are shown in the middle column of Figure 8, and those with normal resolution in left column for comparison. And then use defocused images and displacement to compute the flat-field with KLL algorithm and conduct subsequent analysis. The results in the ROI, including standard deviation, mean value, accuracy of the derived flat-field and possible accuracy of SDI flat-field calibration, are presented in Table 1. Also given in the table is the result without defocusing (the case with defocused resolution of 1.2"). The results indicate that defocusing images (reducing spatial resolution) can largely improve the accuracy of the derived flat-field, and images with worse defocused resolution give better flat-field accuracy. Unfortunately, the resultant accuracy with DF-SAT-KLL method without considering the effect of solar rotation does not meet the requirement of 2%.

As mentioned above, the time span to acquire the 21 off-pointing images is 80 minutes. Tests show that influence of the solar rotation in such a time span on determining the displacement of different images cannot be ignored. To deal with this issue, instead of limb-fitting, we use the local correlation tracking (LCT) algorithm to determine the relative displacement between different images so as to eliminate the influence of the rotation of the Sun. With the displacements between different images and the off-pointing images with different defocused resolutions, we computed the flat-field again via the KLL algorithm and conducted similar analysis to the limb-fitting case.

Figure 6 shows one example of derived flat-field images in the 304 Å and 1700 Å wavebands using off-pointing images with a defocused resolution of 18" and LCT algorithm, and Figure 7 shows the ratio of derived flat-field to the corresponding base flat-field in the ROI. Table 2 shows the statistical results of the derived flat-field images. Comparing the results in Tables 2 and 1, we see that the accuracy of derived flat-field is further improved after taking into the influence of solar rotation. When the defocused resolution is worse than 12", the estimated accuracy of SDI flat-field calibration meets the requirement of 2%.

Table 1: Results with the DF-SAT-KLL Method With the Influence of Solar Rotation.

| Defocused Resolution (″) | 1.2 | | 6.0 | | 12 | | 18 | |
|---|---|---|---|---|---|---|---|---|
| Data Waveband (Å) | 304 | 1700 | 304 | 1700 | 304 | 1700 | 304 | 1700 |
| Mean Value | 1.0767 | 0.9611 | 1.0762 | 0.9584 | 1.0768 | 0.9588 | 1.0780 | 0.9590 |
| Standard Deviation Value | 0.0559 | 0.0401 | 0.0387 | 0.0198 | 0.0297 | 0.0166 | 0.0253 | 0.0164 |
| Flat-field Accuracy (%) | 5.19 | 4.17 | 3.60 | 2.07 | 2.76 | 1.73 | 2.35 | 1.72 |
| SDI's Flat-field Accuracy (%) | 4.68 | | 2.83 | | 2.24 | | 2.03 | |

Table 2: Results with the DF-SAT-KLL Method Without the Influence of Solar Rotation.

| Defocused Resolution (″) | 1.2 | | 9.0 | | 12 | | 15 | | 18 | |
|---|---|---|---|---|---|---|---|---|---|---|
| Data Waveband (Å) | 304 | 1700 | 304 | 1700 | 304 | 1700 | 304 | 1700 | 304 | 1700 |
| Mean Value | 1.0748 | 0.9660 | 1.0738 | 0.9659 | 1.0743 | 0.9659 | 1.0748 | 0.9658 | 1.0753 | 0.9658 |
| Standard Deviation Value | 0.0503 | 0.0390 | 0.0282 | 0.0164 | 0.0251 | 0.0154 | 0.0228 | 0.0148 | 0.0210 | 0.0144 |
| Flat-field Accuracy (%) | 4.68 | 4.04 | 2.63 | 1.69 | 2.34 | 1.59 | 2.12 | 1.53 | 1.96 | 1.49 |
| SDI's Flat-field Accuracy (%) | 4.36 | | 2.16 | | 1.96 | | 1.82 | | 1.72 | |

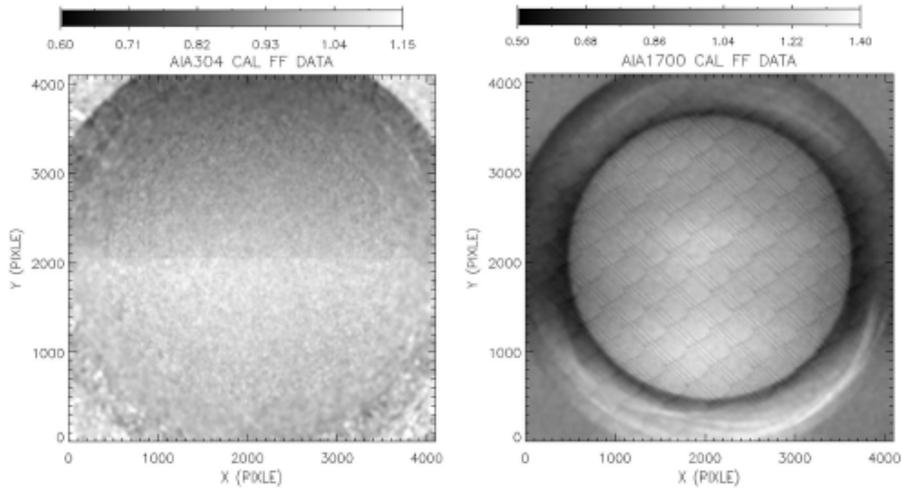

Fig. 6: Derived flat-field AIA in the 304 Å (left) and 1700 Å (right) wavebands using DF-SAT-KLL method and off-pointing images with a defocused resolution of 18" together with LCT for eliminating the influence of solar rotation.

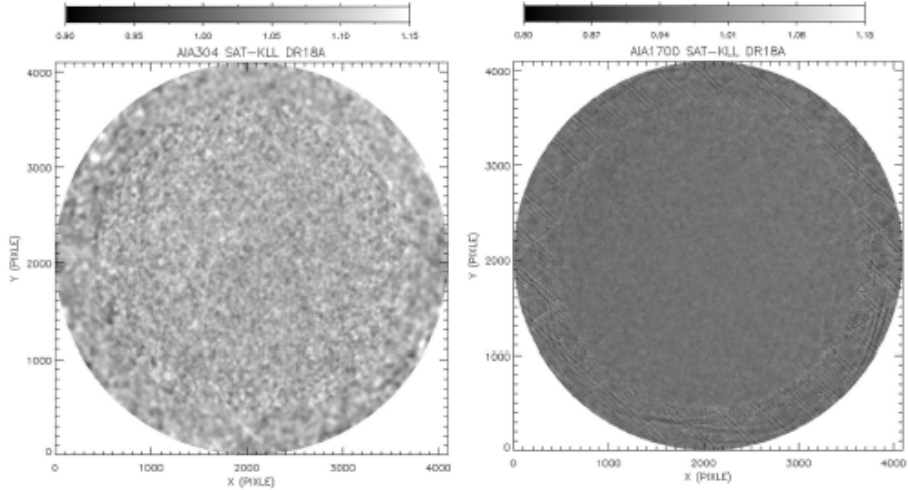

Fig. 7: Ratio of the derived flat-field with DF-SAT-KLL and LCT algorithm to the corresponding base flat-field in the ROI in the 304 Å (left) and 1700 Å (right) wavebands.

Table 3: The Flat-field Accuracy of the DFM-SAT-KLL Method with defocused resolution **12** arc seconds.

| Magnification Factor | 1.2 | | 1.5 | | 2.0 | |
|---|---|---|---|---|---|---|
| Data Waveband (Å) | 304 | 1700 | 304 | 1700 | 304 | 1700 |
| Mean Value | 1.0759 | 0.9660 | 1.0769 | 0.9661 | 1.0766 | 0.9661 |
| Standard Deviation Value | 0.0226 | 0.0146 | 0.0212 | 0.0144 | 0.0220 | 0.0144 |
| Flat-field Accuracy (%) | 2.10 | 1.51 | 1.97 | 1.49 | 2.04 | 1.49 |
| SDI's Flat-field Accuracy (%) | 1.81 | | 1.73 | | 1.77 | |

### 4.3 KLL Method with Image Defocusing and Magnification

Defocused is usually accompanied by magnification of the image of the Sun, which is not included in the DF-SAT-KLL method described in subsection 4.2. In this subsection, we give the results with this effect of image magnification included. We call it DFM-SAT-KLL method for short when image magnification is included.

For the DFM-SAT-KLL method, the off-pointing images required for the KLL algorithm are generated in the following procedures: (1) defocus the downloaded AIA images with a 2D Gaussian PSF as described above and magnify them by a certain factor; (2) shift the defocused and magnified images ac-cording to the 21 offset values; (3) multiply the central part with 4096x4096 pixels (constraint from AIA base flat-field) of the shifted images by the corresponding AIA base flat-field to create the off-pointing images. The defocused images for the reference position from AIA observations with 12" resolution and with magnification of 1.5 times are shown in the right column of Figure 8.

KLL algorithm and LCT algorithm are applied to these images to compute the flat-field. Table 3 shows the results with magnification factor of 1.2, 1.5 and 2.0. It can be seen that the DFM-SAT-KLL method with an appropriate magnification factor can further improve the accuracy of the derived flat-field.

## 5. CALIBRATION FOR THE SCI INSTRUMENT

SCI includes two coronal imaging channels that share optical elements. We have proposed to use a diffuser to conduct the in-flight flat-field calibration of SCI. Chen et al. (2019) has introduced the SCI in-flight flat-field calibration, and we refer readers to the paper for more information. For the entity of this paper, we briefly describe the method. To conduct the in-flight flat-field calibration, a transmissive diffuser is inserted between the primary mirror and the secondary mirror, which is far from the focal plane of the first mirror and therefore is illuminated by the Sun almost uniformly. The transmissive diffuser can work as either a uniform or stable light source for flat-field calibration, depending on the uniformity of the transmitted beam. If the transmitted beam is uniform enough, it can be used as a uniform light source; otherwise, it can be used as a stable light source, because the change of solar irradiance can be ignored in the time span of a few minutes for SCI in-flight flat-field calibration. The calibration accuracy of this method is expected to be 2% (Chen et al. 2019) or better, which meets the in-flight calibration requirement.

## 6. DISCUSSIONS AND CONCLUSIONS

In this paper, we briefly introduce the in-flight flat-field calibration methods for the WST and SCI instruments of the LST payload. We present in much more details the methodology for that of the SDI instrument, taking the so-called KLL algorithm (Kuhn, Lin, Loranz 1991) as a basis, which was first proposed for flat-field calibration of ground-based solar telescope observing the full-disk Sun in the visible waveband. As expected, the Sun is an ideal light source for WST in-flight flat-field calibration and we show that the KLL method is applicable for WST in-flight flat-field calibration, and the accuracy of the derived flat-field is not sensitive to the time interval of two adjacent images. This is because WST images the full-disk photosphere (low atmosphere) in UV white-light (around 3600 Å), which are almost stable in the time span of image acquisition for the flat-field calibration.

SCI is a Lyot type coronagraph working in both visible light and the Lyα wavebands, which adopts similar methods for flat-field calibration to other coronagraphs, such as LASCO, and COR1 and COR2 of the SECCHI suite on SOHO. Namely, SCI employs a transmissive diffuser to provide uniform or stable light sources required for in-flight flat-field calibration. This method has been proven feasible (Brueckner et al. 1995; Howard et al. 2008) and can yield flat-field with high accuracy (Chen et al. 2019).

SDI takes images of the solar lower atmosphere in the Lyα waveband, which shows constant and obvious changes. The necessary images for SDI in-flight flat-field calibration using KLL algorithm are obtained by off-pointing the spacecraft platform, and it requires about 80 minutes for all the images to be record-ed. These images are not suitable for the KLL algorithm because in such a long time interval of about 80 minutes the Sun in the Ly☐ waveband change significantly and the rotation of the Sun itself makes the limb-fitting method for determining the relative displacement not appropriate. This has been shown by our analysis in subsection 4.1.1 (the SAT-KLL method).

We proposed to deal with the change of the Sun by defocusing the off-pointing images and reduce the corresponding spatial resolution, which was achieved by applying 2D Gaussian PSF of certain FWHM to the obtained images in our analysis. This procedure can be done in-flight by inserting additional optical component, say a lens. We also used the LCT algorithm instead of the limb-fitting to determine the relative displacements between different off-pointing images so as to reduce or eliminate the effect of the rotation of the Sun. We have demonstrated that defocusing the images and using LCT both can improve the accuracy of the resultant flat-field. When the defocused resolution is worse than 12" we can obtain the flat-field of SDI with KLL algorithm that satisfies the requirement of 2% (the DF-SAT-KLL method). We could recommend the defocused resolution of 18" in practice bearing in mind that some unexpected factors and uncertainty may affect the results. Our results also show that defocusing the image more can further improve the accuracy of the computed flat-field.

As mentioned above, image defocusing generally happens with image magnification. Appropriate magnification can once more improve the accuracy of the resultant flat-field as shown in subsection 4.3 (DFMSAT-KLL method). When the image of the Sun is magnified, the number of photons received per pixel of the image sensor in unit time decreases. Assuming that the magnification factor is n, then the number of photons received by each pixel in the same time is $1/n^2$ of that without magnification. In order to achieve the same signal noise ratio, the exposure time needs to be increased by a factor of $n^2$. Therefore, the magnification factor should be as small as possible if the defocused resolution meets the requirement.

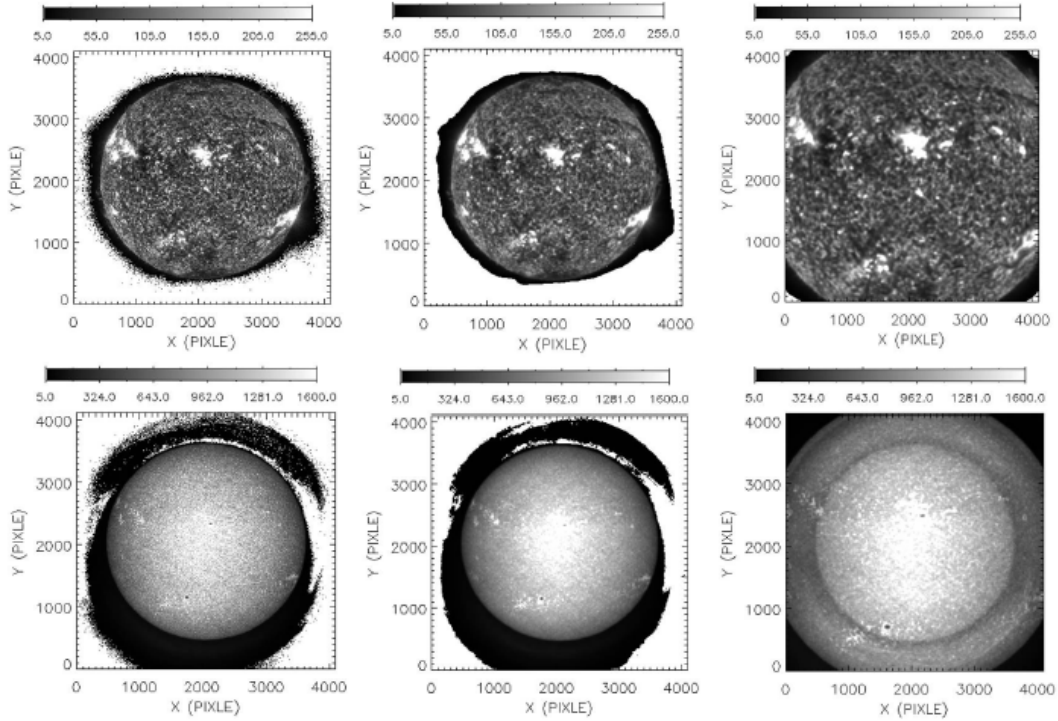

Fig. 8: Generated offset images with normal spatial resolution of 1.2" (left), defocused resolution of 12" (middle), and defocused resolution of 12" and magnification of 1.5 times (right) for the reference position of the 21 off-pointing positions from AIA observations in the 304 Å (upper) and 1700 Å (lower) wavebands.

SDI has also Piezoelectric Transducer (PZT) activators implemented for image stabilization, which can shift the solar image by a small amount of about 50". Therefore, in principle, it can be used to obtain images with small displacement for KLL algorithm in a short time span. Due to the limited displacement that PZT can make, it is impossible to reproduce the structures larger than the shift that PZT can do with KLL algorithm. The base AIA flat-field images (Figure 3) have apparent large-scale structures, which cannot be used to test the KLL method with images obtained using PZT to realize the image displacement.

The signal-to-noise ratio (SNR) is very important and may have significant influence on the obtained flat-field. The SNR of AIA data we used is about 10 and 30 for 304 Å and 1700 Å wavebands, respectively. We also used AIA data obtained in 2017 to simulate the input images for the KLL algorithm and calculate the flat-field. The results showed that low SNR led to worse flat-field accuracy. For LST flat-field calibration, we can set appropriate exposure time to guarantee the obtained images have high enough SNR (at lease >10).

In summary, we have tested the methods for the in-flight flat-field calibrations of the three instruments (WST, SCI and SDI) of the LST payload using data from AIA and HMI observations for SDI and WST, respectively. Normal KLL method is proposed for WST and implementing a transmissive diffuser is adopted for SCI. For the in-flight flat-field calibration of SDI, we recommend the KLL method with off-pointing images with defocused resolution of around 18" (DF-SAT-KLL method), and the LCT algorithm instead of limb-fitting should use to determine the relative displacements between different images.

## ACKNOWLEDGMENTS

This work is supported by the National Natural Science Foundation of China (Grant Nos. U1731241, 11503089 and 11973012) and by the CAS Strategic Pioneer Program on Space Science (Grant Nos. XDA15052200, XDA15320103 and XDA15320301).